\documentclass[aoas]{imsart}

\RequirePackage{amsthm,amsmath,amsfonts,amssymb}
\RequirePackage[authoryear]{natbib}
\RequirePackage[colorlinks,citecolor=blue,urlcolor=blue]{hyperref}
\RequirePackage{graphicx}

\startlocaldefs

\endlocaldefs

\begin{document}

\begin{frontmatter}
\title{A Bayesian hierarchical model for methane emission source apportionment}
\runtitle{Methane emission source apportionment}

\begin{aug}
\author[A]{\fnms{William S.}~\snm{Daniels}\ead[label=e1]{wdanie16@jhu.edu}\orcid{0000-0001-8752-5536}},
\author[A]{\fnms{Douglas W.}~\snm{Nychka}{}\orcid{0000-0003-1387-3356}},
\\ \and
\author[A]{\fnms{Dorit M.}~\snm{Hammerling}{}\orcid{0000-0003-3583-3611}}
\address[A]{Department of Applied Mathematics and Statistics, Colorado School of Mines\printead[presep={,\ }]{e1}}
\end{aug}

\begin{abstract}
Reducing methane emissions from the oil and gas sector is a key component of short-term climate action. Emission reduction efforts are often conducted at the individual site-level, where being able to apportion emissions between a finite number of potentially emitting equipment is necessary for leak detection and repair as well as regulatory reporting of annualized emissions. We present a hierarchical Bayesian model, referred to as the multisource detection, localization, and quantification (MDLQ) model, for performing source apportionment on oil and gas sites using methane measurements from point sensor networks. The MDLQ model accounts for autocorrelation in the sensor data and enforces sparsity in the emission rate estimates via a spike-and-slab prior, as oil and gas equipment often emit intermittently. We use the MDLQ model to apportion methane emissions on an experimental oil and gas site designed to release methane in known quantities, providing a means of model evaluation. Data from this experiment are unique in their size (i.e., the number of controlled releases) and in their close approximation of emission characteristics on real oil and gas sites. As such, this study provides a baseline level of apportionment accuracy that can be expected when using point sensor networks on operational sites.
\end{abstract}


\end{frontmatter}


\section{Introduction} 

By 2050, climate change is projected to cause an additional 14.5 million deaths and \$12.5 trillion in economic losses worldwide due to an increase in the frequency and severity of extreme weather events \citep{climate_change}. Rising concentrations of carbon dioxide and methane in the atmosphere are the primary drivers of climate change, making it crucial to reduce emissions of these gasses \citep{IPCC2021_WGI_SPM}. Methane, in particular, provides an important opportunity to mitigate the near-term impacts of climate change given its relatively short lifetime in the atmosphere and high heat trapping potential compared to carbon dioxide \citep{IPCC}.

The oil and gas industry accounts for 21\% of global anthropogenic methane emissions \citep{iea2025} and is a viable opportunity for emissions reductions given that: 1) emissions from equipment malfunctions can often be mitigated quickly once they are identified, and 2) improvements in technology and facility design can eliminate emissions from normally operating equipment such as pneumatic controllers \citep{nisbet_methane_2020}. Knowing which pieces of equipment are emitting and their emission rates over time is necessary to both prioritize mitigation efforts and verify that long term emission reduction strategies are working. However, estimating emission source and rate based on ambient methane measurements is a challenging inverse problem due to the turbulent nature of near-field atmospheric transport, the degree of variability in emission rates, and the fact that only a subset of oil and gas equipment are emitting at a given time.

In this article, we analyze in situ sensor data from an experimental oil and gas site where methane is released at known rates in a way that emulates real-world emission characteristics. This site is called the Methane Emissions Technology Evaluation Center (METEC). Sensor data collected at METEC come as close as possible to data from a real oil and gas site while still knowing ground truth emission source and rate. We develop a Bayesian hierarchical model to estimate emission source and rate at METEC, and by extension, on real oil and gas sites. The model accommodates multisource scenarios and assumes that potential source locations are known a priori. We call this model the multisource detection, localization, and quantification (MDLQ) model. This is the first study, to our knowledge, that pairs a novel method for source apportionment with a controlled release experiment that is sufficiently realistic to generalize to operating oil and gas sites.


\subsection{Methane releases at METEC} \label{sec:intro_metec}

\begin{figure}[t]
\centering
\includegraphics[width=0.86\linewidth]{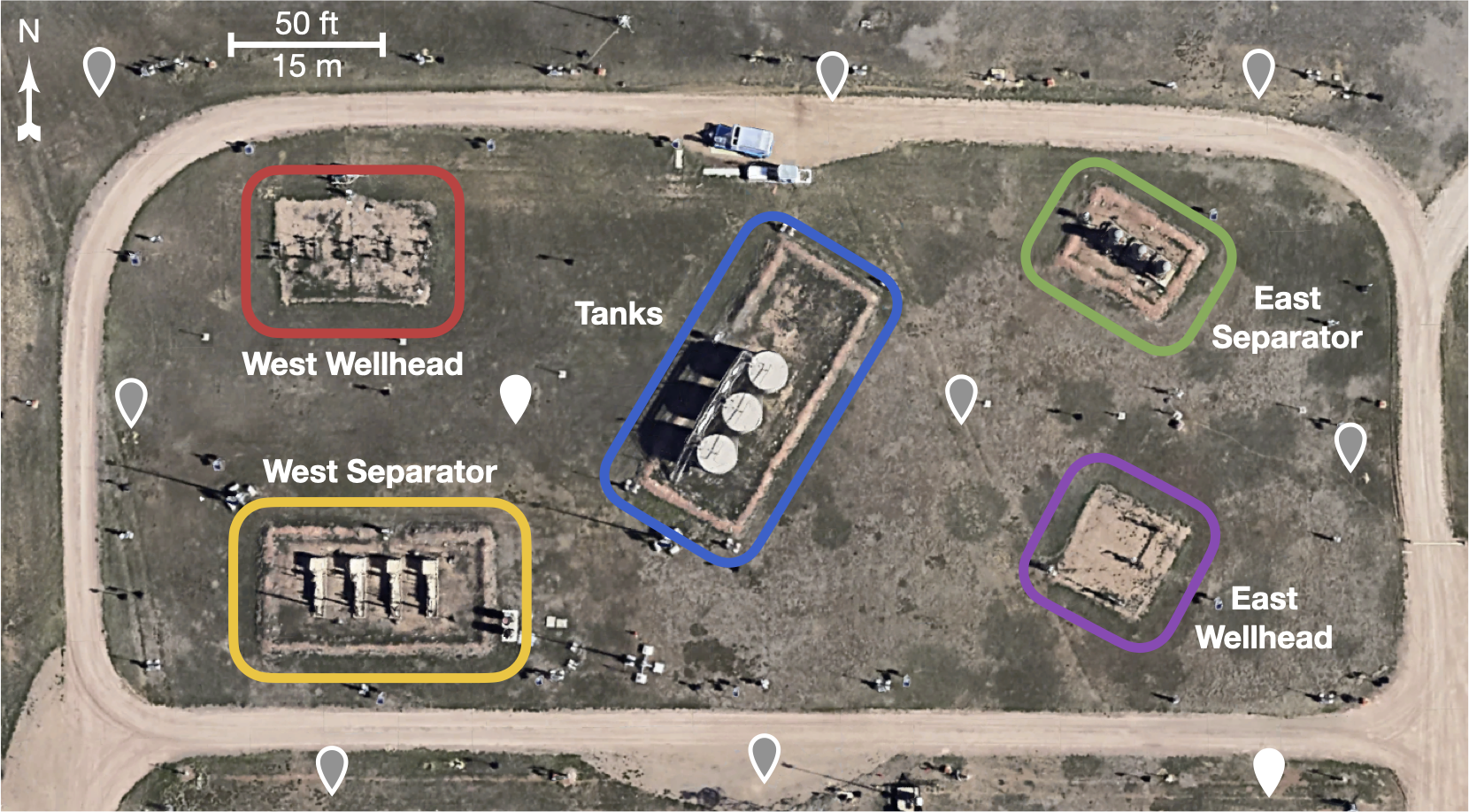}
\vspace{-0.25cm}
\caption{\textit{Aerial view of the METEC facility located in Fort Collins, Colorado. The five potential emission sources are identified with colored squares, and the locations of the 10 methane sensors are marked with pins. Pins with a gray interior show sensors that measure wind speed and direction in addition to methane concentrations.}}
\label{fig:METEC}
\end{figure}

The METEC facility (shown in Figure \ref{fig:METEC}) is instrumented with real oil and gas equipment, such as wellheads and separators, but does not actually produce oil or gas. Instead, known quantities of methane are released from typical leak points on the equipment. From February 5 to April 30, 2024, 337 controlled releases were conducted at METEC in a way that emulates the following characteristics of real oil and gas methane emissions: emissions can last from minutes to days with rates that vary over time \citep{daniels_estimating_2024}, any subset of the equipment can emit simultaneously \citep{wang_multiscale_2022}, and there can be periods of no emissions \citep{cusworth_intermittency_2021}.

This experiment provides a much larger and more realistic sample of methane emissions than other site-level inversion studies in the literature. For example, \cite{cartwright_bayesian_2019} analyze two releases, \cite{newman_probabilistic_2024} analyze three releases, and \cite{kumar_near-field_2022} analyze 26 releases. The ability to apportion methane emissions using a small number of sensors is influenced by site layout, wind, and other meteorological conditions, making it important to assess the performance of a statistical inversion over a wide range of conditions. 


Furthermore, the releases in this study better emulate the emission characteristics of active oil and gas sites than similar studies in the literature. For example, \citet{weidmann_locating_2022} evaluate their method on methane releases with known start and end times, and \citet{hirst_methane_2020} perform releases in flat, empty locations with simple atmospheric transport. As such, the modeling results in this study are more likely to generalize to real oil and gas sites of similar size and complexity to METEC and provide a baseline level of apportionment accuracy on these sites.

\subsection{The continuous monitoring inverse problem}\label{sec:intro_inversion}

The MDLQ model uses methane concentration measurements from a network of in situ sensors placed directly on oil and gas sites to apportion emissions. We refer to this type of sensor network as a continuous monitoring system (CMS). The locations of the CMS sensors in this study are shown as pins in Figure \ref{fig:METEC}. Each sensor measures ambient methane concentrations once per second, which are averaged in situ to the minute-frequency before being stored and used for all subsequent analysis. This averaging step reduces data volume while still capturing short-lived enhancements at the sub-minute scale. 

Let $\boldsymbol{y}_k \in \mathbb{R}^l$ denote the minute-frequency concentration data from sensor $k$, where $l$ is the number of minutes within the temporal domain of the inversion (discussed in Section \ref{sec:moving_window}). The full observation vector is then defined as $\boldsymbol{y} = (\boldsymbol{y}_1^T, ..., \boldsymbol{y}_m^T)^T \in \mathbb{R}^{n}$, where $m$ is the number of CMS sensors installed on the site and $n = l \times m$ is the total number of observations across sensors. As with most remote sensing technologies, an atmospheric inversion is required to translate $\boldsymbol{y}$ into estimates of emission source and rate.


Many of the existing inversion techniques for CMS rely on the assumption that one source is emitting at a time \citep{xue_turbulent_2017, cartwright_bayesian_2019, kumar_near-field_2022, daniels_detection_2024}, which is a known simplification of real emission characteristics \citep{wang_multiscale_2022, kunkel_extension_2023}. Inversion techniques that allow for multiple simultaneous emission sources often assume that the CMS concentration measurements are a linear combination of concentration predictions from an atmospheric transport model, 
\begin{equation}\label{eqn:data_level}
    \boldsymbol{y} = X \boldsymbol{\beta} + \boldsymbol{\epsilon},
\end{equation}

\noindent where emission rates for the $p$ potential sources on the site, $\boldsymbol{\beta} = (\beta_1, ..., \beta_p)^T \in \mathbb{R}^p$, are assumed to be constant within the temporal domain of the inversion. The design matrix is often constructed as $X = [\boldsymbol{x}^{(1)}\cdot\cdot\cdot \boldsymbol{x}^{(p)}] \in \mathbb{R}^{n \times p}$, where each $\boldsymbol{x}^{(i)} \in \mathbb{R}^n$ for $i=\{1,...,p\}$ contains simulated methane concentrations from the transport model at all sensor locations assuming that source $i$ is emitting at a unitary rate. Similar to $\boldsymbol{y}$, each $\boldsymbol{x}^{(i)}$ concatenates the $l$ simulated concentrations at each of the $m$ sensors into one vector. In many studies, the errors $\boldsymbol{\epsilon} = (\epsilon_1, ..., \epsilon_n)^T \in \mathbb{R}^n$ are assumed normal, independent, and identically distributed \citep{ars_statistical_2017,  cartwright_bayesian_2019, hirst_methane_2020, newman_probabilistic_2024}, which is another limitation given that $\boldsymbol{y}$ are time series data. In this study, we are primarily concerned with estimates for $\boldsymbol{\beta}$, as these tell us the subset of sources that are emitting and their rate. 

Our approach, implemented in the MDLQ model, assumes the same linear model as Equation \ref{eqn:data_level} at the data level, but further models $\boldsymbol{\beta}$ as a mixture distribution with a point mass at zero. This prior model allows for identically zero emission rate estimates, a necessary property given that potential sources are often not emitting, and extends the method in \cite{weidmann_locating_2022}, who used a narrow Gaussian to approximate the point mass. This extension eliminates the need to tune the relative widths of the components of the mixture distribution. Finally, following \cite{ganesan_characterization_2014}, we allow $\boldsymbol{\epsilon}$ to be autocorrelated since the elements of $\boldsymbol{y}$ are observed over time at 1-minute intervals. 


\subsection{Outline}

In Section \ref{sec:data}, we provide additional details about the METEC experiment, atmospheric transport model, and our background-removal procedure. In Section \ref{sec:model}, we present the MDLQ model and our procedure for running it operationally, rather than on experiments with known start and end times. In Section \ref{sec:results}, we present source apportionment results for the METEC experiment, probe the limitations of the MDLQ model via a simulation study, and compare our results to current state-of-the-art methods from the literature. 

\newpage
\section{Data} \label{sec:data}

\subsection{METEC controlled release and measurement data} \label{sec:controlled_release}

Methane concentration data for the METEC experiment come from a network of 10 CMS sensors installed around the facility at a height of 2 m. Eight of the sensors are also equipped with an anemometer to measure wind speed and direction in the horizontal plane. Wind data are necessary to simulate the transport of methane from the sources to the sensors (discussed further in Section \ref{sec:dispersion_model}). The properties (e.g., accuracy and precision) of the CMS sensors in this study are comparable to the sensors being used operationally to mitigate emissions. Full sensor specifications are provided in Section S1 of the Supporting Information (SI) file.

Unlike many studies in the literature, the controlled releases in this experiment were performed around-the-clock to evaluate the model under both daytime and nighttime meteorology. Each release consists of emissions from a subset of the five potential emission sources shown in Figure \ref{fig:METEC}. Emission rates were constant within a given release, but could be different between the emitting equipment. Across the 337 releases, equipment-level release rates ranged from 0.08 to 7.2 kg/hr, emission durations ranged from 1.1 minutes to 8.0 hours, and source heights ranged from 0.91 m (West Wellheads) to 5.0 m (Tanks).

For this study, the exact source locations and heights were provided by the METEC operations team. In practice, these quantities must be estimated before performing source apportionment with the MDLQ model. Publicly available satellite imagery can be used for an accurate estimate of location (e.g., latitude and longitude) and a coarse estimate of height. More precise estimates of source height can be obtained directly from the oil and gas operator or through detailed drone imagery. For pipelines, source locations can be assigned at maintenance access points and control valves, where a large majority of pipeline emissions originate \citep{zimmerle_methane_2015}. 

\begin{figure}[t]
\centering
\includegraphics[width=\linewidth]{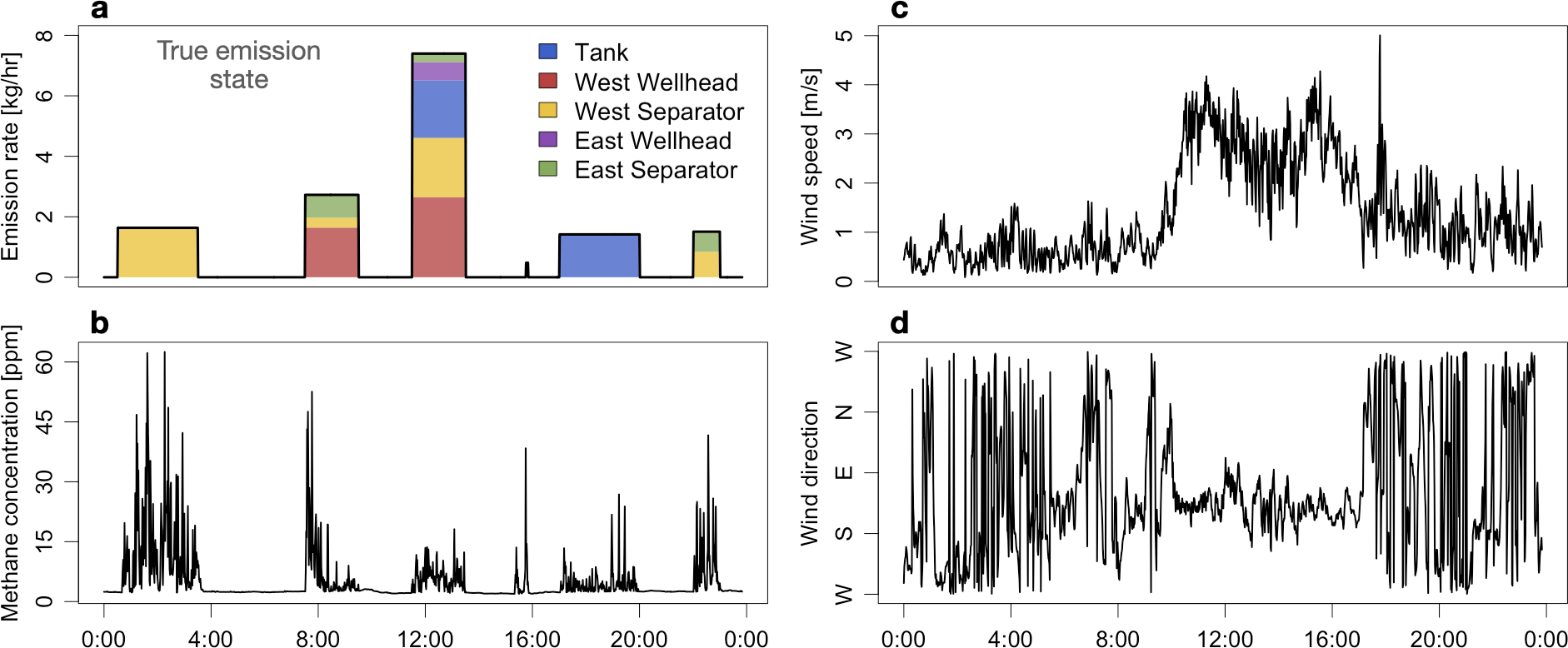}
\vspace{-0.75cm}
\caption{\textit{Example data from February 12, 2024. (a) True release data, with color corresponding to the emission source. (b) Minute-by-minute maximum of the concentration observations across the 10 CMS sensors. (c) Minute-by-minute median of the wind speed observations across the 8 sensors with anemometers. (d) Minute-by-minute circular median of the wind direction observations across the 8 sensors with anemometers.}}
\label{fig:data_example}
\end{figure}

As an example of the METEC controlled release data, Figure \ref{fig:data_example} shows the true emission state during a 24-hour period and the corresponding concentration and wind data from the CMS sensors. In Figure \ref{fig:data_example}(b), we plot the minute-by-minute maximum concentration observation across the 10 CMS sensors as a condensed summary of the data. Figure S4 in the SI shows the methane concentration measurements from all 10 CMS sensors separately. Notice that the concentration enhancements align with the timing of the controlled releases, but the amplitude of the enhancements do not follow the magnitude of the releases. This is because the sensors are not located at the release points, and meteorological effects such as wind speed and dispersion affect the transport of methane and the resulting concentrations at the sensor locations. Modeling this transport is critical for an accurate inversion of emission source and rate.

\subsection{Atmospheric transport model} \label{sec:dispersion_model}

We use a Gaussian puff model (GPM) to simulate the transport of methane from the emission sources to the sensor locations. The GPM approximates a continuous release of methane by simulating the movement of many small ``puffs'' of methane that are modeled as 3-dimensional Gaussian-like distributions. Discretizing a continuous release of methane in this manner allows the GPM to move each puff according to the wind conditions at its location in both space and time, providing a more accurate representation of atmospheric transport than models that assume steady state conditions. Note that the GPM treats methane as a passive tracer in the atmosphere, which is a reasonable assumption at the time and distance scales used for source apportionment on oil and gas sites. We describe the GPM below, but see \citet{jia_fast_2025} or \citet{stockie_mathematics_2011} for more details.

Consider a coordinate system in which emission source $i$ is located at $(0,0,H_i)$. Puffs of methane from this source will all originate at this point. To avoid confusion with the symbols in Equation \ref{eqn:data_level}, we refer to locations in this coordinate system as $(x', y', z')$. New puffs are created at a given frequency, $\delta t$, by default once per second, and the coordinates are rotated for each puff so that the positive $x'$-direction is pointed in the downwind direction at the time of puff creation. In this coordinate system, the predicted methane concentration at sensor location $(x', y', z')$ from a given puff, index by $j$, that originated at $(0,0,H_i)$ and has been in existence for time $t$ is modeled as
\begin{equation}
\begin{split}
    c_j^{(i)}(x',y',z',t) =& \frac{Q_j}{(2\pi)^{3/2}  \sigma_y^2 \sigma_z} \exp{\left(-\frac{(x'-u_j t)^2+y'^2}{2\sigma_y^2}\right)} \\
    &\left[\exp{\left(-\frac{(z'-H_i)^2}{2\sigma_z^2}\right)} +
    \exp{\left(-\frac{(z'+H_i)^2}{2\sigma_z^2}\right)} \right],
\end{split}
\end{equation}

\noindent where $Q_j$ is the mass of methane contained in puff $j$, $\sigma_y$ and $\sigma_z$ are the dispersion parameters in the $y'$ and $z'$ directions, respectively, and $u_j$ is the wind speed at the time of puff creation. The value of $Q_j$ is a function of the specified emission rate, $q$, and the puff creation frequency, $\delta t$. The GPM requires time, $t$, to be discretized at a fixed multiple of the puff creation frequency, $\delta t$.

At each subsequent time step, a new puff is created at $(0,0,H_i)$ and puffs already in existence are transported based on the wind speed and direction from their respective creation times. The GPM used in this study assumes zero advection in the vertical direction, as wind data in this dimension are hard to obtain in practice, but the puffs do grow more diffuse in all three physical dimension as they move downwind. The total methane concentration at sensor location $(x', y', z')$ and time $t$ from source $i$ is then  
\begin{equation}
    c^{(i)}(x', y', z', t) = \sum_{j=1}^P c_j^{(i)}(x', y', z', t),
\end{equation}

\noindent where $P$ is the number of puffs in existence at time $t$. After running the GPM, we average $c^{(i)}$ to the minute-frequency to align with the sensor data. The simulations are conducted at a higher frequency to capture the second-frequency transport dynamics that are also captured in the CMS observations.

The dispersion parameters $\sigma_y$ and $\sigma_z$ control the size of each puff as it moves downwind and becomes more diffuse over time. We specify $\sigma_y$ and $\sigma_z$ using the EPA parameterization of the Pasquill-Gifford-Turner dispersion scheme \citep{Pasquill1961, Turner1970, EPA1992}. Under this parameterization, both $\sigma_y$ and $\sigma_z$ are functions of stability class and monotonically increase with total distance traveled (see Section S2 in the SI for details). 

\begin{figure}[t]
\centering
\includegraphics[width=\linewidth]{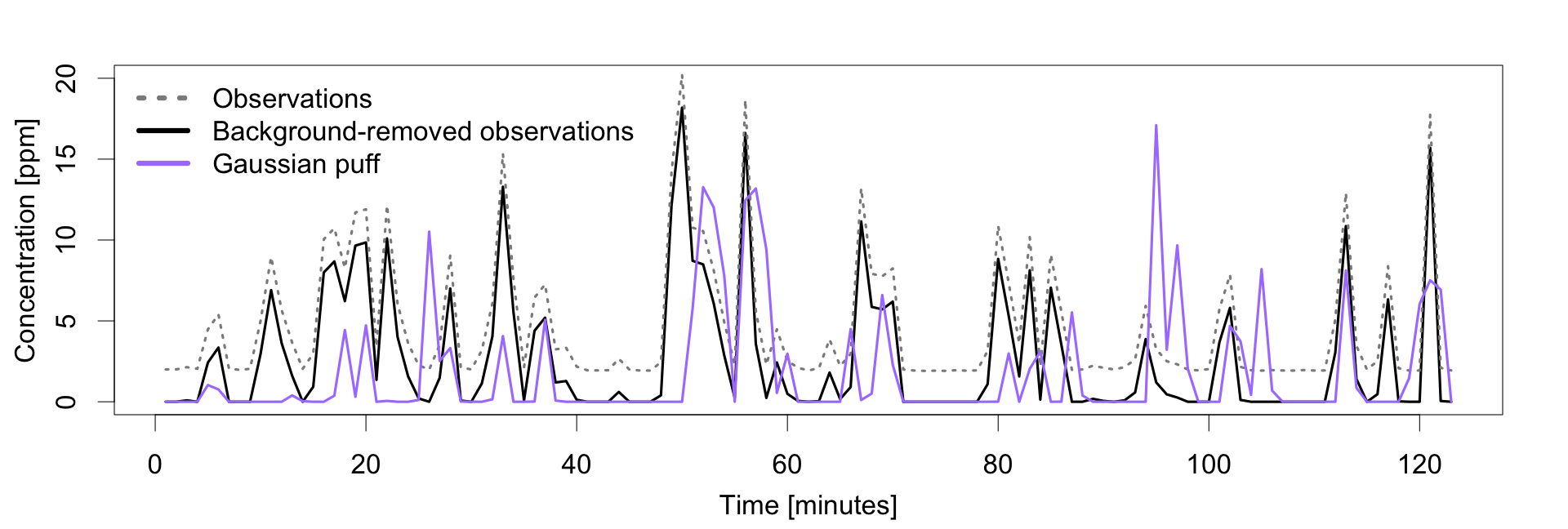}
\vspace{-0.75cm}
\caption{\textit{Methane concentration observations and concentration predictions from the Gaussian puff atmospheric dispersion model during one of the METEC controlled releases.}}
\label{fig:concentration_example}
\end{figure}

As an example, Figure \ref{fig:concentration_example} shows concentration observations from one of the CMS sensors during a controlled release and the corresponding output from the GPM scaled by the true release rate. Background-removed concentration observations are also included in this figure, which we discuss next in Section \ref{sec:background}.

\subsection{Background estimation and removal} \label{sec:background}

Background methane concentrations are always present in the atmosphere but are not modeled by the GPM. We must therefore estimate the methane background and remove it from the CMS observations before evaluating model-data mismatch, as failing to do so would introduce upward bias in the emission rate estimates from the MDLQ.

\citet{hirst_methane_2020} include background concentrations as a parameter to be determined in their inversion model. However, we find that the methane background can be accurately estimated before running an inversion by using the gradient-based spike detection algorithm from \citet{daniels_detection_2024}. This algorithm identifies sharply elevated concentration observations (``spikes'') in the CMS data and estimates a local background for each as an average of the concentration observations immediately preceding and following the spike. Figure \ref{fig:concentration_example} shows the raw concentration observations from one of the CMS sensors and the background-corrected concentrations using this algorithm.

This method for estimating the methane background does not require fitting a full spatiotemporal model. Furthermore, we have found that estimating the background using a fixed value, such as the 5$^\text{th}$ percentile from \citet{cartwright_bayesian_2019}, often underestimates the true background by a few parts per million. This can lead to large errors in the inferred emission rates when the amplitude of the concentration enhancements is on a similar scale as the errors in the background estimate.

\section{Model}\label{sec:model}

\subsection{Bayesian hierarchical model} \label{sec:model_def}

\begin{figure}[t]
\centering
\includegraphics[width=0.75\linewidth]{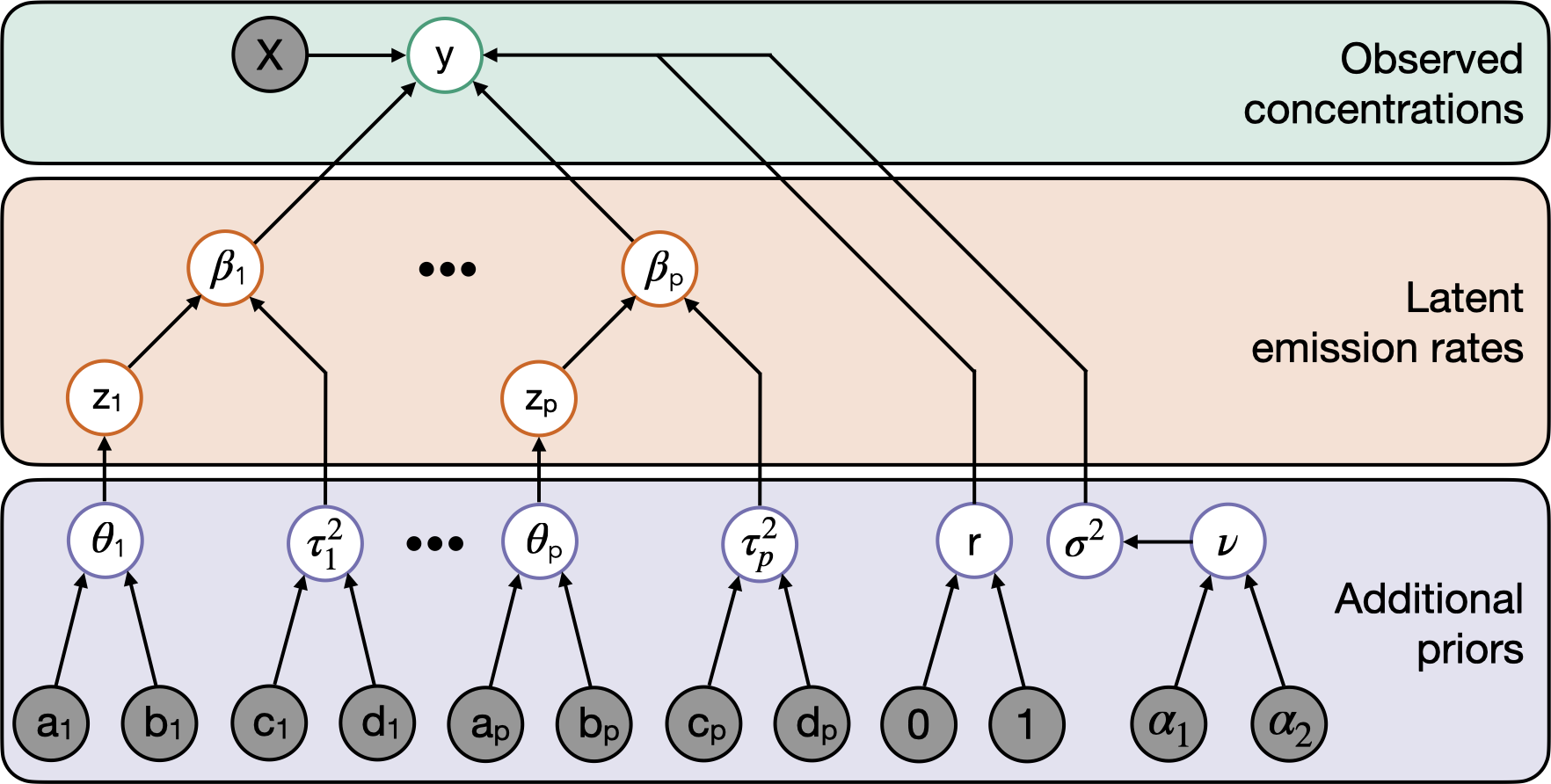}
\caption{\textit{Hierarchical structure and conditional independencies of the MDLQ model. Each circle represents a parameter, with fixed hyperparameters in gray and random variables in white. Arrows indicate when one parameter is a function of other parameters.}}
\label{fig:DAG}
\end{figure}

The MDLQ model has a hierarchical structure, summarized in Figure \ref{fig:DAG}. The highest level contains a model for the observed methane concentrations. The next level contains a sparsity-inducing prior for the unobserved methane emission rates. The lowest level contains priors for all other model parameters that are not of direct interest for methane emission source apportionment but still contain important uncertainty information. 

We model the CMS concentration observations as a linear combination of concentration predictions from the GPM, 
\begin{equation}\label{eqn:MDLQ_data_level}
\boldsymbol{y} = X \boldsymbol{\beta} + \boldsymbol{\epsilon},
\end{equation}

\noindent where emission rates for the $p$ potential sources on the site, $\boldsymbol{\beta} = (\beta_1, ..., \beta_p)^T \in \mathbb{R}^p$, are assumed to be constant within the temporal domain of the inversion, $\mathcal{D} = \{1,...,l\}$. To account for the fact that oil and gas methane emissions vary over time, we apply the MDLQ model to sequential, non-overlapping domains of length $l=30$ minutes as new CMS data become available. This windowed approach is described fully in Section \ref{sec:moving_window}; the remainder of this section defines the MDLQ model for one such temporal domain.

As in Section \ref{sec:intro_inversion}, we denote the background-removed, minute-frequency concentration observations from sensor $k$ as $\boldsymbol{y}_k \in \mathbb{R}^l$. We then define the full observation vector as $\boldsymbol{y} = (\boldsymbol{y}_1^T, ..., \boldsymbol{y}_m^T)^T \in \mathbb{R}^{n}$, where $m$ is the number of CMS sensors installed on the site and $n = l \times m$ is the total number of observations across sensors. This formulation allows us to leverage information from all sensors jointly instead of performing a separate inversion for each. 

Using the notation from Section \ref{sec:dispersion_model}, we denote GPM predictions at sensor $k$ as 
\begin{equation}
    \boldsymbol{x}^{(i)}_k= (c^{(i)}(x_k', y_k', z_k', 1), ..., c^{(i)}(x_k', y_k', z_k', l))^T \in \mathbb{R}^l,
\end{equation}
\noindent assuming that source $i$ is emitting at a unitary rate ($q$ = 1 g/s). Note that $\boldsymbol{x}^{(i)}_k$, and hence $\boldsymbol{\beta}$, are a linear function of $q$. However, we scale the posterior estimate of $\boldsymbol{\beta}$ by $q$, so this GPM parameter does not impact the source apportionment results. Similar to the definition of $\boldsymbol{y}$, we concatenate GPM predictions at all sensor locations into a univariate vector given by 
\begin{equation}
\boldsymbol{x}^{(i)} = ({\boldsymbol{x}^{(i)}_1}^T, ..., {\boldsymbol{x}^{(i)}_m}^T)^T \in \mathbb{R}^n.
\end{equation}
\noindent Finally, we define the design matrix as 
\begin{equation}
X = [\boldsymbol{x}^{(1)}\cdot\cdot\cdot \boldsymbol{x}^{(p)}] \in \mathbb{R}^{n \times p}.
\end{equation}

We assume that the errors $\boldsymbol{\epsilon} = (\epsilon_1, ..., \epsilon_n)^T \in \mathbb{R}^n$ follow a multivariate normal, such that $\boldsymbol{\epsilon} \sim N(0, \sigma^2 R)$, where $R \in \mathbb{R}^{n \times n}$ is an AR(1) correlation matrix with entries $[R]_{i,j}=r^{|i-j|}$ and $r \in [0,1)$ is the autoregressive coefficient or correlation parameter. Under this parameterization, $\sigma^2$ represents the marginal error variance rather than the innovation variance of the AR(1) process. Modeling autocorrelation in $\boldsymbol{\epsilon}$ extends previous work and results in better coverage of the credible intervals for $\boldsymbol{\beta}$, shown via a simulation study in Section S8 of the SI. 

The order of the autoregressive error model was selected via a visual analysis of partial autocorrelation functions (PACFs), shown in Section S11 of the Supporting Information. Specifically, we ran the MDLQ model on the METEC data using an AR(1) error model, computed residuals $\hat{\boldsymbol{\epsilon}} = \boldsymbol{y} - X\hat{\boldsymbol{\beta}}$, and formed the implied one-step-ahead innovations $\hat{\boldsymbol{\eta}}_t = \hat{\boldsymbol{\epsilon}}_t - \hat{r}\hat{\boldsymbol{\epsilon}}_{t-1}$. We then examined PACFs of $\hat{\boldsymbol{\eta}}_t$ within each inversion window, noting that the innovations should be uncorrelated under a correctly specified AR(1) model. While modest residual autocorrelation remains, we do not believe that it warrants a higher autoregressive order, especially since the additional complexity could lead to identifiability issues with $\boldsymbol{\beta}$.

The emission rate parameter for each source, $\boldsymbol{\beta}_i$ takes the form of a mixture model with a point mass at 0 kg/hr, given by
\begin{align}
\beta_i &\sim 
   \begin{cases} 
      0 & z_i = 0 \\
      \text{Exp}(\tau_i^2) & z_i = 1,
   \end{cases}
\end{align}

\noindent where the indicator $z_i \in \{0,1\}$ is modeled as $z_i \sim \text{Bernoulli}(\theta_i)$. This model is an example of a sparsity enforcing spike-and-slab prior \citep{mitchell_bayesian_1988}. Sparsity in the emission rates is an important property of the MDLQ model, as emission sources on oil and gas sites are often not emitting. Previous work has found that emission rates over time follow right skewed distributions \citep{daniels_toward_2023, khaliukova_investigating_2025}, so we use an exponential for the non-zero component of the mixture parameterized by scale $\tau_i^2$. The exponential also forces $\boldsymbol{\beta}$ to have non-negative support, which is a necessary constraint given that there are no methane sinks on oil and gas sites. 

This variant of the spike-and-slab prior uses a discrete point-mass as the spike rather than a continuous distribution, such as an exponential with scale parameter much smaller than $\tau_i^2$. This allows for $\beta_i$ to be sampled as identically zero, rather than a very small value. Furthermore, it eliminates the need to prescribe or infer the scale parameter of the narrow distribution approximating the discrete point-mass, which can be challenging to do in such a way that the MCMC sampler is efficient and mixes well.

The remaining model parameters are given the following priors:
\begin{align}
   \theta_i &\sim \text{Beta}(a_i, b_i)\\
   \tau_i^2 &\sim \text{Inv-Gamma}(c_i, d_i)\\
   \sigma^2 &\sim \text{Inv-Gamma}(\nu/2, \nu/2)\\
   \nu &\sim \text{Inv-Gamma}(\alpha_1, \alpha_2)\\
   r &\sim \text{Uniform}(0,1). 
\end{align}

Fixed hyperparameters $a_i$, $b_i$, $c_i$, and $d_i$ control any prior information about the probability that $\beta_i$ is non-zero and the scale of $\beta_i$. For example, heaters on oil and gas sites are not used in the summer, and as such the $a_i$ and $b_i$ hyperparameters could be set such that $\theta_i$ has more mass close to zero in the prior during these months. In this article, however, we use a uniform prior for $\theta_i$ by setting $a_i = b_i =1$ and relatively uninformative priors for the remaining parameters by setting $c_i = d_i = \alpha_1 = \alpha_2 = 1$. These values were selected to reflect prior knowledge that oil and gas methane emissions (on all sites, not just METEC) are mostly concentrated at small values between 0 and 10 kg/hr, with rare emissions occurring on the order of 100 to 100,000 kg/hr. In particular, setting $c_i=d_i=1$ results in an Inverse Gamma prior for $\tau^2_i$ with undefined mean and variance; this allows for very large $\tau^2_i$ to be sampled without putting too much mass at extreme, unrealistic values. Emissions estimates from the MDLQ model are not highly sensitive to the choice of $c_i$ and $d_i$ for similarly realistic values. See SI Section S12 for more details and a sensitivity study of these prior hyperparameters.

The prior for $\sigma^2$ was set such that, in the prior model, the $\boldsymbol{\epsilon}$ are distributed according to a Student's t-distribution with $\nu$ degrees of freedom, as we discovered that $\boldsymbol{\epsilon}$ can be heavier tailed than a Gaussian through basic model diagnostics. This prior belief can be updated by the data. We find that emission rate estimates from the MDLQ model are not sensitive to the choice of $\alpha_1$ and $\alpha_2$; see SI Section S12 for details.

\subsection{Posterior inference} \label{sec:inference}

We sample from the joint posterior of all model parameters using a Markov chain Monte Carlo (MCMC) algorithm. We describe our MCMC algorithm below, but see \citet{gelman_bayesian_2015} for a more complete discussion of Bayesian inference via MCMC. While samples are obtained for all model parameters, the emission rates, $\boldsymbol{\beta}$, and indicator variables, $\boldsymbol{z}$, are the most useful parameters for methane emission source apportionment on oil and gas sites.


We use a Metropolis-within-Gibbs algorithm to sample from the posterior. We are able to derive a closed form expression for the conditional distributions of all model parameters except $\nu$ and $r$, resulting in a fast sampler that mixes well. For $\nu$ and $r$, we update using two Metropolis-Hastings steps within the Gibbs sampler that use a Gaussian as the candidate distribution. We use the posterior mean of the MCMC samples as a point estimate of emission rate in all subsequent analysis. Section S3 in the SI contains a derivation of each conditional distribution used in the sampler.

\subsection{Inference on a moving time window} \label{sec:moving_window}


The MDLQ model assumes that emission rates are constant over the temporal domain of the inversion, $\mathcal{D}=\{1,...,l\}$ even though methane emissions on oil and gas sites are known to vary over time. As such, this assumption breaks down for large $l$. To address this issue, we partition the full time series of concentration observations into consecutive, non-overlapping windows of length $l = 30$ minutes and treat each window as an independent inverse problem. For a given window, $\boldsymbol{y} \in \mathbb{R}^n$ denotes the concatenated minute-frequency concentration observations across all sensors within that 30-minute period, exactly as defined in Section \ref{sec:model_def}, with $n = l \times m$. Likewise, the corresponding design matrix $X \in \mathbb{R}^{n \times p}$ is constructed from the GPM simulations for the same 30-minute period. For the METEC experiment, with $l = 30$, $m = 10$, and $p=5$, we get $\boldsymbol{y} \in \mathbb{R}^{300}$ and $X \in \mathbb{R}^{300\times 5}$ for each window.

This windowed approach yields a piecewise-constant representation of emission behavior over time, with temporal resolution determined by the window length, $l$. The choice of a 30-minute window reflects a tradeoff between model assumptions and stability of the inferred parameters. Shorter windows are more likely to satisfy the constant emission assumption but provide fewer observations, leading to higher uncertainty and increased sensitivity to transport model errors. Longer windows provide more data for each inversion but are increasingly likely to violate the constant emission rate assumption. Based on exploratory analysis and a formal sensitivity study presented in Section S4 of the SI, we find that a 30-minute window provides a practical compromise for the METEC dataset.

\subsection{Assessing the information content of each inversion window}\label{sec:convergence}

Occasionally, there are no CMS sensors downwind of one or more sources for an entire 30-minute inversion window. When this happens, the sensors will not detect elevated concentration values if one of these sources is emitting, and as such, they provide no information about the emission state of these sources. We therefore avoid making inference on sources with no downwind sensors during a given inversion window. We can identify these sources before running the inversion via the design matrix, $X$; columns of $X$ that consist of only zero concentration predictions likely have no downwind sensor during that inversion window. 

In this study, we say that the ``viable sources'' for a given window are those whose columns of $X$ contain at least four non-zero concentration predictions. We intentionally chose a low threshold so that weakly informative inversion windows are still included in our analysis. This allows us to gauge how incomplete sensor coverage influences source apportionment accuracy and is discussed further in Section \ref{sec:metec_results} and SI Section S5. For each window, we subset $X$ and $\boldsymbol{\beta}$ to just the viable sources and run the inversion as described in Section \ref{sec:inference}, resulting in a rate estimate for only the viable sources. The percent of inversion windows with information on each subset of sources is given in Table \ref{tab:info}.

\begin{table}[t]
\centering
\caption{Information content of the CMS data across the METEC experiment.}
\vspace{0.25cm}
\begin{tabular}{|c|c|c|c|c|c|c|}
\hline
\textbf{\begin{tabular}[c]{@{}c@{}}Number of sources\\ with downwind sensors\end{tabular}} & 5      & 4      & 3     & 2     & 1     & 0     \\ \hline
\textbf{\begin{tabular}[c]{@{}c@{}}Percent of \\ inversion windows\end{tabular}}           & 69.4\% & 17.5\% & 8.7\% & 3.9\% & 0.2\% & 0.2\% \\ \hline
\end{tabular}
\label{tab:info}
\end{table}

We note that $\boldsymbol{y}$ will be identically $\boldsymbol{0}$ for some inversion windows (e.g., from 4:00 to 8:00 in Figure \ref{fig:data_example}). This corresponds to periods where the CMS sensors did not detect methane concentrations above background levels. When this occurs, we set the emission rate for all viable sources to 0 kg/hr without running the MDLQ model. As discussed above, we make no inference on sources with no downwind sensors, and as such, we do not set their emission rate to zero in this situation.  

\subsection{Emissions inventories and alerts}\label{sec:inventories}

Emission mitigation using CMS is done through two primary mechanisms: emissions inventories and alerts. Below, we define these terms and describe how they are constructed using MDLQ output.

A source-level inventory estimates the total emitted methane from a given source over a specific period of time (here, the duration of the METEC study). Inventories are used to implement regulatory penalties for excess emissions and to evaluate long-term emission reduction strategies. We construct source-level inventories by multiplying the posterior mean of $\boldsymbol{\beta}$ in each inversion window by the length of the window and then summing across the experiment. We quantify uncertainty by sampling from the posterior of $\boldsymbol{\beta}$ in each window, constructing a source-level inventory for each sample, and reporting the $0.025^\text{th}$ and $0.975^\text{th}$ percentiles as a 95\% credible interval on the inventory. We construct a site-level inventory by summing the source-level inventories. For windows where a source has no emissions estimate due to a lack of information (i.e., no downwind sensor), we sample from that source's emissions estimates in other windows. This approach is justified by the low correlation between emission rate and wind direction, which determines when a source has downwind sensors.

Emission alerts provide a point estimate of which sources are emitting during a given inversion window. They are used by oil and gas operators to determine whether detected emissions are from normally operating equipment or from malfunctioning equipment that require immediate intervention. The posterior probability that source $i$ is emitting is given by the proportion of samples with $z_i = 1$. We therefore classify source $i$ as emitting in a given window if the posterior mean of $z_i$ exceeds its average across all other windows. We do not create alerts when the CMS data do not contain information for a given source.

\section{Results} \label{sec:results}

\subsection{Methane emission source apportionment at METEC} \label{sec:metec_results}

\begin{figure}[t]
\centering
\includegraphics[width=0.9\linewidth]{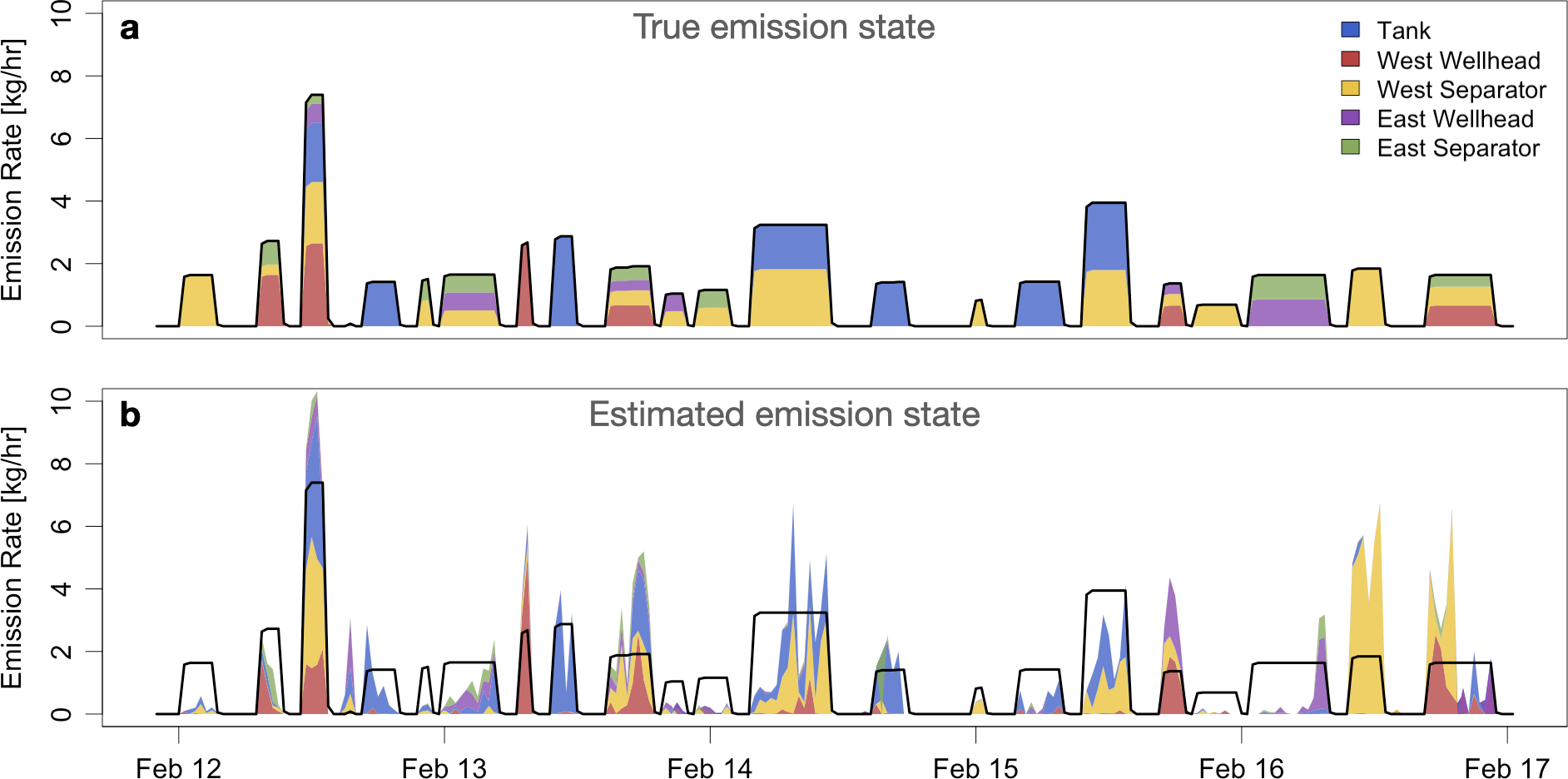}
\vspace{-0.25cm}
\caption{\textit{Output from the MDLQ model during 5 days of the METEC controlled releases. (a) shows the true emission state and (b) shows the estimated emission state from the MDLQ model. Color corresponds to the emitting equipment, and the black line shows the true site-level total emission rate over time.}}
\label{fig:result_time_series}
\end{figure}

We begin by showing output from the MDLQ model over an example 5-day period in Figure \ref{fig:result_time_series}. We make two observations. First, the MDLQ model successfully distinguishes periods of emissions from periods of no emissions; it rarely infers a non-zero emission rate when no controlled releases are occurring, and vice versa (see Table \ref{tab:detection} for details). Second, inferred emission rates vary over the course of each release, despite the true emission rates being constant. This is because the 30-minute inversion windows are shorter than most releases, and inadequacies of the GPM will often cause individual inversion windows to either over or underestimate the true emission rate. We discuss MDLQ model refinements that may mitigate this behavior in Section \ref{sec:discussion}.

\begin{figure}[t]
\centering
\includegraphics[width=\linewidth]{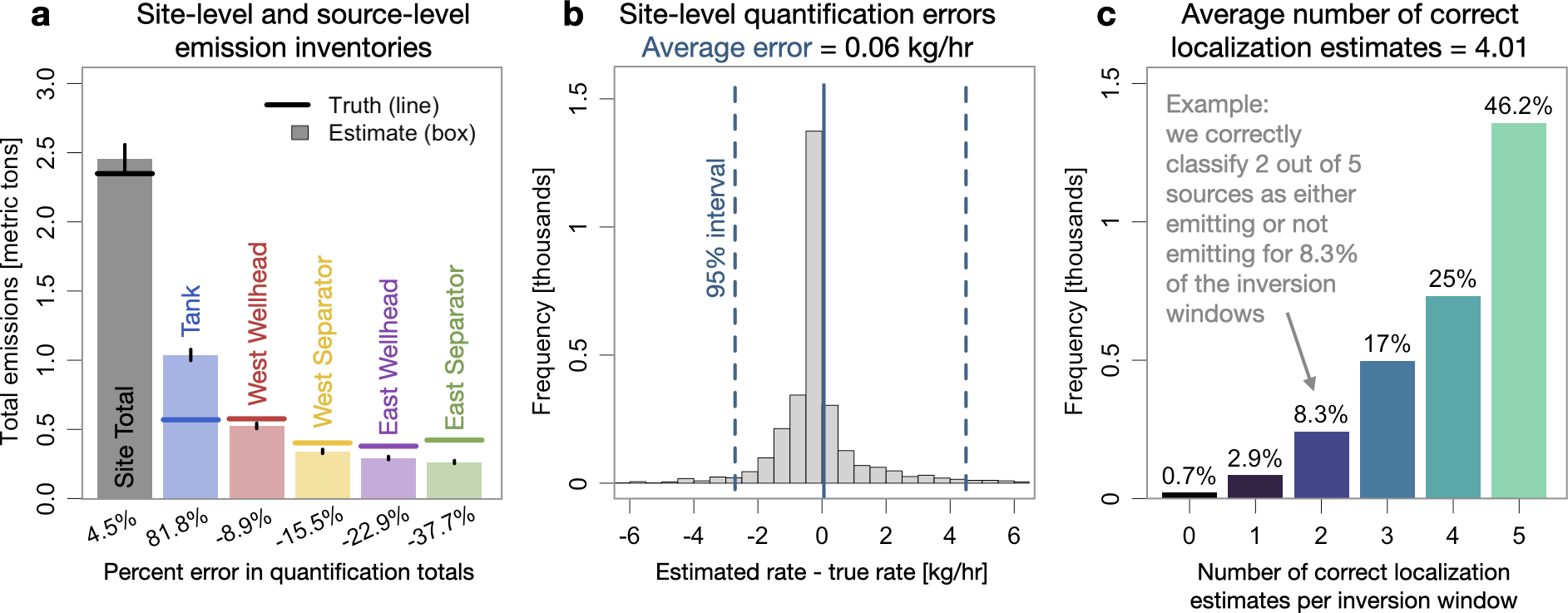}
\vspace{-0.75cm}
\caption{\textit{Summary of the source apportionment results for the METEC controlled releases. (a) Site- and source-level inventories created by summing the emission rate estimates across all inversion windows. Vertical lines show 95\% credible intervals. (b) Distribution of errors in the site-level emission rate estimates. (c) Alerting accuracy, where a correct localization estimate is a correct estimate of emission state (i.e., either emitting or not emitting).}}
\label{fig:quantification_results}
\end{figure}

We now summarize our source apportionment results over the entire METEC experiment, focusing on the MDLQ model's ability to produce inventories and alerts. Figure \ref{fig:quantification_results}(a) shows site- and source-level inventories for the study period. The site-level inventory is close to the truth, overestimating by only 4.5\%. This is promising, as site-level inventories are the primary tool for emissions accounting in US regulations \citep{us_environmental_protection_agency_greenhouse_2024}, EU import standards \citep{EUreg}, and voluntary emission management protocols \citep{OGMP}. However, the tank inventory is overestimated by 81.8\% and the east separator inventory is underestimated by -37.7\%. The remaining source-level inventories are slightly underestimated with errors ranging from -8.9\% to -22.9\%. The cause of these source-level inventory errors is investigated fully in Section S5 of the SI, but in short, they are likely a result of inversion windows with downwind sensors for only a subset of the window. When there is no downwind sensor for the entire window, the MDLQ model does not infer an emission rate (as discussed in Section \ref{sec:convergence}). However, when only a subset of the window has a downwind sensor, it naively appears as if no emissions occurred during the remainder of the window and an emission rate is still inferred. This happens more often for the sources that are close to the perimeter of the site and less often for the tanks, which are centrally located. As a result of this geometry, the MDLQ model likely overestimates tank emissions to compensate for the missed emissions from the remaining sources. To test this, we keep only the inversion windows that have downwind sensors for the entire window and re-calculate the inventories. The resulting site-level inventory remains accurate (error = 2.0\%) and the accuracy of the source-level inventories are much improved, with errors ranging from -22.0\% to 20.4\% (much closer to the inventories' credible intervals). This filtering, however, significantly reduces the amount of data available to apportion emissions.

Figure \ref{fig:quantification_results}(b) shows the distribution of errors in the site-level emission rate estimates across all inversion windows. The average error is 0.06 kg/hr, which makes sense given the small error in the site-level inventory from Figure \ref{fig:quantification_results}(a). However, the inner 95\% interval of the site-level errors is [-2.7, 4.5] kg/hr, meaning that there is relatively large variability in the individual site-level rate estimates. We also observed this fact in the example in Figure \ref{fig:result_time_series}. This finding gives rise to the question: how many emission rate estimates need to be averaged before the site-level estimate is close to the truth? We investigate this question by resampling from the site-level error distribution at different sample sizes (see SI Section S6); we find that 95\% of the site-level sample means are within [-1, 1] kg/hr after averaging just 34 inversion windows (or approximately 17 hours). The average emissions estimate converges quickly because there are no large outliers in the site-level error distribution for the METEC experiment.

Although this fact is not shown in Figure \ref{fig:quantification_results}(b), we note that the coverage of the 95\% credible intervals for the site-level rate estimates is only 0.66, meaning that we have missed a source of uncertainty in the MDLQ model. The missing uncertainty is very likely coming from bias in the GPM, as this model (while more accurate than comparable dispersion models) is still an oversimplification of actual atmospheric transport. We test this theory via a simulation study discussed shortly in Section \ref{sec:simulation_study}.

Figure \ref{fig:quantification_results}(c) assesses the accuracy of the alerts generated by the MDLQ model. For each inversion window, we check if the localization estimate (i.e., the emission state estimate) for each source is correct. That number, ranging from 0 to 5, is plotted in Figure \ref{fig:quantification_results} for all inversion windows. On average, the MDLQ correctly identifies 4.01 out of the 5 possible source as either emitting or not emitting. It most frequently identifies all 5 sources correctly, doing so for 46.2\% of the windows. Table \ref{tab:detection} shows additional detection metrics at both the source- and site-level. The MDLQ model detects 83.8\% of emissions at the site-level, but is only able to detect 56.2\% to 67.9\% of emissions at the source-level. In other words, if a piece of equipment is emitting, the MDLQ model will likely detect it and generate an alert for the site. However, determining exactly which subset of sources is emitting is a harder problem, and the MDLQ model is more prone to errors in this use case. Full source- and site-level confusion matrices are provided in Section S7 of the SI.

\begin{table}[h]
\caption{Source- and site-level detection metrics for the METEC experiment. TPR = true positive rate, TNG = true negative rate, PPV = positive predictive value, and NPV = negative predictive value.}
\begin{tabular}{c|c|c|c|c|c|c|}
\cline{2-7}
                                        & \textbf{\begin{tabular}[c]{@{}c@{}}West\\ Wellhead\end{tabular}} & \textbf{\begin{tabular}[c]{@{}c@{}}West\\ Separator\end{tabular}} & \textbf{Tanks} & \textbf{\begin{tabular}[c]{@{}c@{}}East\\ Wellhead\end{tabular}} & \textbf{\begin{tabular}[c]{@{}c@{}}East\\ Separator\end{tabular}} & \textbf{\begin{tabular}[c]{@{}c@{}}Site-\\ level\end{tabular}} \\ \hline
\multicolumn{1}{|c|}{\textbf{TPR}}      & 67.9\%                                                           & 56.2\%                                                            & 60.6\%         & 64.4\%                                                           & 61.2\%                                                            & 83.8\%                                                         \\ \hline
\multicolumn{1}{|c|}{\textbf{TNR}}      & 90.5\%                                                           & 90.4\%                                                            & 77.4\%         & 90.7\%                                                           & 92.3\%                                                            & 91.2\%                                                         \\ \hline
\multicolumn{1}{|c|}{\textbf{PPV}}      & 73.7\%                                                           & 73.6\%                                                            & 55.0\%         & 74.5\%                                                           & 80.6\%                                                            & 95.1\%                                                         \\ \hline
\multicolumn{1}{|c|}{\textbf{NPV}}      & 87.8\%                                                           & 81.3\%                                                            & 81.1\%         & 85.8\%                                                           & 82.0\%                                                            & 73.4\%                                                         \\ \hline
\multicolumn{1}{|c|}{\textbf{Accuracy}} & 84.1\%                                                           & 79.4\%                                                            & 72.1\%         & 82.9\%                                                           & 81.7\%                                                            & 86.2\%                                                         \\ \hline
\end{tabular}
\label{tab:detection}
\end{table}

\subsection{Simulation study} \label{sec:simulation_study}

We conduct a simulation study to isolate the effect of known biases in the GPM from other potential sources of error. Specifically, we study the impact of misaligned concentration enhancements between the GPM predictions and CMS sensor observations, examples of which can be seen in Figure \ref{fig:concentration_example}. This behavior is a result of various inadequacies of the GPM, such as not resolving turbulent eddy motion or obstructions.

\begin{figure}[b]
\centering
\includegraphics[width=\linewidth]{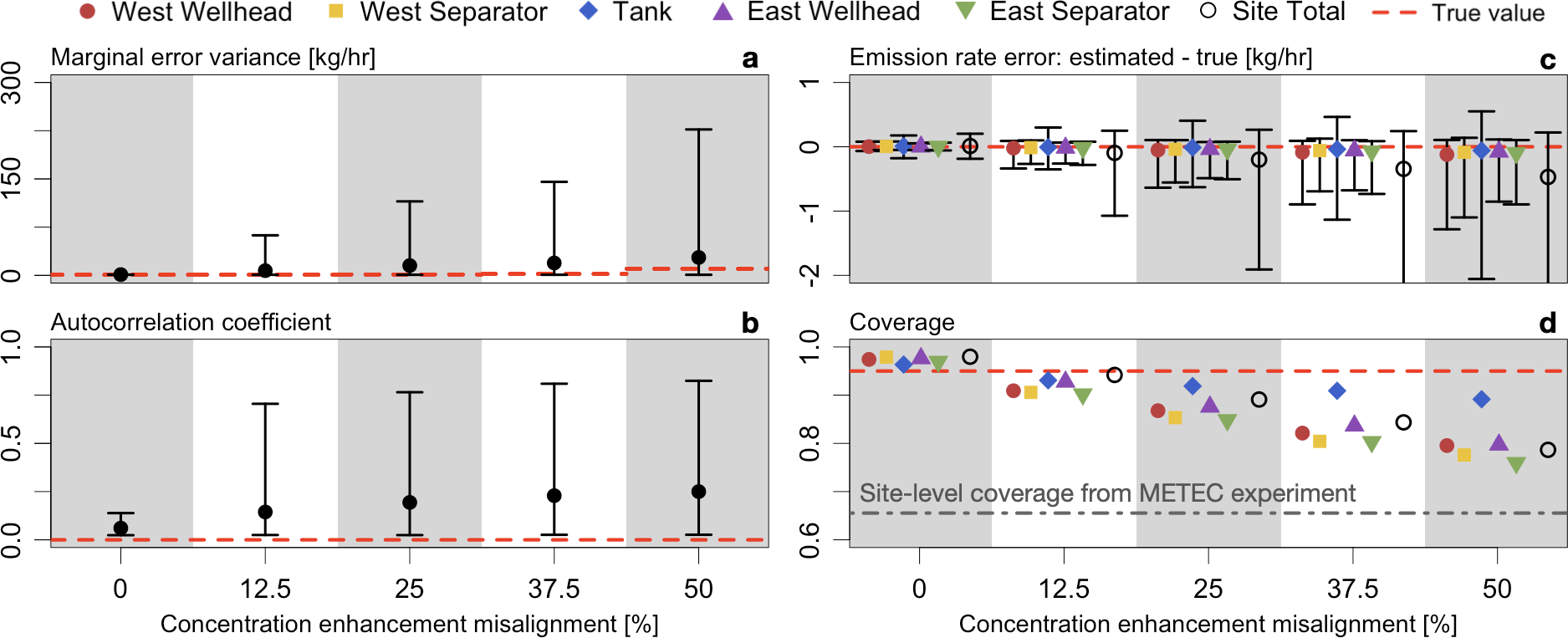}
\vspace{-0.5cm}
\caption{\textit{Simulation study showing MDLQ performance under different levels of forward model bias. Red horizontal lines show the true values for each response feature. Bars show the inner 95\% range for each value across the inversion windows. Colored shapes correspond to source-level results, and black circles correspond to site-level results.}}
\label{fig:simulation}
\end{figure}

We design the simulation study using the same physical geometry (i.e., source and sensor locations), wind data, and emission characteristics (i.e., timing and rate) as the METEC experiment. For each 30-minute inversion window, we construct the artificial observations as 
\begin{equation}
    \Tilde{\boldsymbol{y}} = X \boldsymbol{\beta}_T + \Tilde{\boldsymbol{\epsilon}},
\end{equation}

\noindent where $\boldsymbol{\beta}_T$ are the true emission rates, $X$ is the GPM output (as defined in Section \ref{sec:model}), and $\Tilde{\boldsymbol{\epsilon}} \sim N(0,1)$. This setup eliminates forward model errors from $\Tilde{\boldsymbol{y}}$, allowing us to reintroduce them in a controlled manner to gauge their impact on the MDLQ output. We reintroduce GPM bias by first identifying groups of concentration enhancements in $\Tilde{\boldsymbol{y}}$ using the gradient-based spike detection algorithm from \citet{daniels_detection_2024}. We then move $M$ percent of the enhancements to a different (uniformly sampled) time during the inversion window. For this study, we test five different values for $M$: \{0, 12.5, 25, 37.5, 50\}. We selected a maximum misalignment of 50\% by examining real model-data mismatch during the METEC experiment; the percent of misaligned concentration enhancements rarely exceeded 50\%. After introducing the artificial GPM bias, we run the MDLQ as described in Section \ref{sec:moving_window} using $\Tilde{\boldsymbol{y}}$ instead of the actual CMS observations.

Figure \ref{fig:simulation} shows the results of this simulation study. We plot four response features for each value of $M$: the marginal error variance, the autocorrelation coefficient, error in the emission rate estimates, and the coverage of the emission rate credible intervals. In subfigures (a)-(c), we plot the mean of each response feature across all inversion windows, with black bars showing their inner 95\% range. In subfigures (c)-(d), we show results separated by emission source and when summed to the site-level. For subfigure (d), coverage is computed as the percent of inversion windows where the true emission rate was contained within the corresponding 95\% credible interval. 

We make three observations. First, all parameters are accurately estimated in the absence of forward model bias, as expected, with source-level coverages close to 95\%. Second, we see a diminishing ability to estimate $\sigma^2$ and $r$ as the degree of forward model bias increases, but emission rate estimates remain largely unbiased with a slight tendency to underestimate. This is promising, as $\boldsymbol{\beta}$ are the parameters of direct interest for methane emission source apportionment. The overestimation of $\sigma^2$ is because the misaligned concentration enhancements result in extremely large errors, either positive or negative, which the Gaussian error model is not flexible enough to accommodate. Some of these errors are likely misinterpreted as autocorrelation, resulting in overestimates of $r$. Finally, source and site-level coverages drop as the degree of forward model bias increases and approach the coverage from the METEC experiment. A drop in coverage is expected, as the misaligned enhancements result in either over or underestimation of $\boldsymbol{\beta}$, such that many of the 95\% credible intervals (which are not aware of the GPM bias) no longer contain the truth. Site-level coverage for the METEC experiment was 0.66, and source-level coverages ranged from 0.71 to 0.81. The fact that coverage values from the simulation study approach those from the METEC experiment provides evidence that GPM bias is indeed a large cause of the underestimated uncertainty in the MDLQ model. The fact that the simulation study coverages are not identical to the real data coverages means that there are additional missing sources of uncertainty beyond misaligned concentration enhancements. 

Section S8 in the SI contains another simulation study demonstrating the importance of modeling autocorrelation in $\boldsymbol{\epsilon}$; failing to do so results in overestimated $\boldsymbol{\beta}$ and low coverages.

\subsection{Comparison to alternative methods}

We analyze the METEC experiment using two other statistical techniques (OLS and LASSO) and three state-of-the-art inversion methods from the literature, with results summarized in Table \ref{tab:comparison}. We select two single-source methods to compare against \citep{cartwright_bayesian_2019, daniels_detection_2024}, both of which assume that only one source is emitting at a time. To apply them to the METEC data, we use the localization procedure from \citet{daniels_detection_2024} to estimate a most likely source for each inversion window before the inversion is performed. It is important to benchmark the performance of single-source methods on multisource releases, despite this being a clear violation of their core assumption, because these methods are still widely used in practice by sensor technology vendors. As such, this comparison provides a baseline level of apportionment accuracy that should be expected on real oil and gas sites from single-source inversion systems. 

We also compare to pyELQ, a multisource inverse model based on the method from \citet{weidmann_locating_2022} and \citet{ijzermans_long-term_2024}. The pyELQ model can estimate methane fluxes on a fixed grid or identify emitting sources using a reversible jump MCMC sampler. To provide a fair comparison, we run pyELQ with a fixed grid that is constrained to the locations of the five sources on the METEC site, as the MDLQ model specifies these sources a priori. Finally, we also compare to a variant of the MDLQ model that assumes independent Laplace errors. Through basic model diagnostics, we found that errors from the MDLQ model often have heavier tails than a Gaussian. These large errors are likely from instances in which a spike in the concentration predictions is not present in the observations, or vice versa. We believed that a Laplace error model might be better able to capture large errors and hence result in better coverage. Section S9 of the SI fully defines this model variant. As none of these alternative methods discuss how to best use their inversion method in practice (rather than on releases with known start and end times), we use the moving time window approach discussed in Section \ref{sec:moving_window} for all methods.

The MDLQ model has the smallest error in the site-level inventory and in the individual site-level emission rate estimates. No one method has the smallest error across all five source-level inventories. The pyELQ model produces the best Tank inventory, LASSO produces the best West Wellhead, West Separator, and East Wellhead inventories, and OLS produces the best East Separator inventory. The MDLQ model variant with Laplace errors has the smallest IQR. The MDLQ model has the best site-level coverage of the 95\% credible intervals. OLS has the best localization performance, followed by the Laplace model variant. Note that for OLS, LASSO, and pyELQ, we assume that a source is emitting if its rate estimate is above 0.1 kg/hr, which was hand-tuned to provide the best possible localization performance on these data.

\begin{table}[t]
\caption{\textit{Comparison to alternative methods using data from the METEC controlled releases. Site-level total error is the error (estimate - truth) in the site-level inventory. The following five rows provide the source-level inventory errors (estimate - truth). Errors are presented in units of metric tons (t) and as percent errors. Average site-level quantification error is the average error (estimate - truth) across all individual site-level emission rate estimates, and the IQR of these errors is the difference between their 25th and 75th percentiles. Coverage of the site-level 95\% credible intervals is the fraction of inversion windows where the true site-level emission rate was contained within the 95\% credible interval of the estimate. The average number of correct localization estimates is the average number of correct state estimates (either on or off) across all inversion windows.}}
\vspace{0.25cm}
\begin{tabular}{|c|cc|cc|ccc|}
\hline
\textbf{}                                                                                            & \multicolumn{2}{c|}{\textbf{\begin{tabular}[c]{@{}c@{}}Other statistical\\ methods\end{tabular}}}                                                             & \multicolumn{2}{c|}{\textbf{\begin{tabular}[c]{@{}c@{}}Single-source\\ methods\end{tabular}}}                                                                                    & \multicolumn{3}{c|}{\textbf{\begin{tabular}[c]{@{}c@{}}Multisource\\ methods\end{tabular}}}                                                                                                                                                                 \\ \hline
\textbf{}                                                                                            & \multicolumn{1}{c|}{\textbf{OLS}}                                                       & \textbf{LASSO}                                                      & \multicolumn{1}{c|}{\textbf{\begin{tabular}[c]{@{}c@{}}Daniels\\ et al.\\ (2024)\end{tabular}}} & \textbf{\begin{tabular}[c]{@{}c@{}}Cartwright\\ et al. \\ (2019)\end{tabular}} & \multicolumn{1}{c|}{\textbf{pyELQ}}                                                     & \multicolumn{1}{c|}{\textbf{\begin{tabular}[c]{@{}c@{}}MDLQ\\ Laplace\\ errors\end{tabular}}} & \textbf{MDLQ}                                                     \\ \hline
\textbf{\begin{tabular}[c]{@{}c@{}}Site-level \\ total error\end{tabular}}                           & \multicolumn{1}{c|}{\begin{tabular}[c]{@{}c@{}}1.32 t\\ (56.2\%)\end{tabular}}          & \begin{tabular}[c]{@{}c@{}}0.60 t\\ (25.6\%)\end{tabular}           & \multicolumn{1}{c|}{\begin{tabular}[c]{@{}c@{}}0.24 t\\ (10.3\%)\end{tabular}}                  & \begin{tabular}[c]{@{}c@{}}1.45 t\\ (61.9\%)\end{tabular}                      & \multicolumn{1}{c|}{\begin{tabular}[c]{@{}c@{}}-0.27 t\\ (-11.6\%)\end{tabular}}        & \multicolumn{1}{c|}{\begin{tabular}[c]{@{}c@{}}1.08 t\\ (46.0\%)\end{tabular}}                & \textbf{\begin{tabular}[c]{@{}c@{}}0.11 t\\ (4.5\%)\end{tabular}} \\ \hline
\textbf{\begin{tabular}[c]{@{}c@{}}West \\ wellhead\\ total error\end{tabular}}                      & \multicolumn{1}{c|}{\begin{tabular}[c]{@{}c@{}}0.15 t\\ (26.0\%)\end{tabular}}          & \textbf{\begin{tabular}[c]{@{}c@{}}0.02 t\\ (2.8\%)\end{tabular}}   & \multicolumn{1}{c|}{\begin{tabular}[c]{@{}c@{}}0.15 t\\ (26.0\%)\end{tabular}}                  & \begin{tabular}[c]{@{}c@{}}-0.02 t\\ (-3.9\%)\end{tabular}                     & \multicolumn{1}{c|}{\begin{tabular}[c]{@{}c@{}}-0.16 t\\ (-27.0\%)\end{tabular}}        & \multicolumn{1}{c|}{\begin{tabular}[c]{@{}c@{}}0.13 t\\ (22.7\%)\end{tabular}}                & \begin{tabular}[c]{@{}c@{}}-0.05 t\\ (-8.9\%)\end{tabular}        \\ \hline
\textbf{\begin{tabular}[c]{@{}c@{}}West\\ separator\\ total error\end{tabular}}                      & \multicolumn{1}{c|}{\begin{tabular}[c]{@{}c@{}}0.07 t\\ (17.2\%)\end{tabular}}          & \textbf{\begin{tabular}[c]{@{}c@{}}-0.02 t\\ (-5.2\%)\end{tabular}} & \multicolumn{1}{c|}{\begin{tabular}[c]{@{}c@{}}-0.05 t\\ (-12.5\%)\end{tabular}}                & \begin{tabular}[c]{@{}c@{}}0.02 t\\ (5.5\%)\end{tabular}                       & \multicolumn{1}{c|}{\begin{tabular}[c]{@{}c@{}}-0.07 t\\ (-17.0\%)\end{tabular}}        & \multicolumn{1}{c|}{\begin{tabular}[c]{@{}c@{}}0.16 t\\ (39.5\%)\end{tabular}}                & \begin{tabular}[c]{@{}c@{}}-0.06 t\\ (-15.5\%)\end{tabular}       \\ \hline
\textbf{\begin{tabular}[c]{@{}c@{}}Tank\\ total error\end{tabular}}                                  & \multicolumn{1}{c|}{\begin{tabular}[c]{@{}c@{}}0.95 t\\ (167.7\%)\end{tabular}}         & \begin{tabular}[c]{@{}c@{}}0.64 t\\ (112.0\%)\end{tabular}          & \multicolumn{1}{c|}{\begin{tabular}[c]{@{}c@{}}0.25 t\\ (44.5\%)\end{tabular}}                  & \begin{tabular}[c]{@{}c@{}}0.31 t\\ (54.1\%)\end{tabular}                      & \multicolumn{1}{c|}{\textbf{\begin{tabular}[c]{@{}c@{}}0.08 t\\ (14.0\%)\end{tabular}}} & \multicolumn{1}{c|}{\begin{tabular}[c]{@{}c@{}}0.39 t\\ (67.7\%)\end{tabular}}                & \begin{tabular}[c]{@{}c@{}}0.47 t\\ (81.8\%)\end{tabular}         \\ \hline
\textbf{\begin{tabular}[c]{@{}c@{}}East\\ wellhead\\ total error\end{tabular}}                       & \multicolumn{1}{c|}{\begin{tabular}[c]{@{}c@{}}0.12 t\\ (32.0\%)\end{tabular}}          & \textbf{\begin{tabular}[c]{@{}c@{}}0.05 t\\ (12.5\%)\end{tabular}}  & \multicolumn{1}{c|}{\begin{tabular}[c]{@{}c@{}}0.05 t\\ (14.4\%)\end{tabular}}                  & \begin{tabular}[c]{@{}c@{}}0.67 t\\ (177.8\%)\end{tabular}                     & \multicolumn{1}{c|}{\begin{tabular}[c]{@{}c@{}}-0.06 t\\ (-15.8\%)\end{tabular}}        & \multicolumn{1}{c|}{\begin{tabular}[c]{@{}c@{}}0.17 t\\ (46.0\%)\end{tabular}}                & \begin{tabular}[c]{@{}c@{}}-0.09\\ (-22.9\%)\end{tabular}         \\ \hline
\textbf{\begin{tabular}[c]{@{}c@{}}East\\ separator\\ total error\end{tabular}}                      & \multicolumn{1}{c|}{\textbf{\begin{tabular}[c]{@{}c@{}}0.03 t\\ (6.1\%)\end{tabular}}}  & \begin{tabular}[c]{@{}c@{}}-0.08 t\\ (-18.7\%)\end{tabular}         & \multicolumn{1}{c|}{\begin{tabular}[c]{@{}c@{}}-0.17 t\\ (-39.4\%)\end{tabular}}                & \begin{tabular}[c]{@{}c@{}}0.47 t\\ (112.0\%)\end{tabular}                     & \multicolumn{1}{c|}{\begin{tabular}[c]{@{}c@{}}-0.07 t\\ (-16.3\%)\end{tabular}}        & \multicolumn{1}{c|}{\begin{tabular}[c]{@{}c@{}}0.23 t\\ (54.7\%)\end{tabular}}                & \begin{tabular}[c]{@{}c@{}}-0.16 t\\ (-37.7\%)\end{tabular}       \\ \hline
\textbf{\begin{tabular}[c]{@{}c@{}}Avg site-level\\ quant error\end{tabular}}                        & \multicolumn{1}{c|}{\begin{tabular}[c]{@{}c@{}}0.55\\ kg/hr\end{tabular}}               & \begin{tabular}[c]{@{}c@{}}0.28\\ kg/hr\end{tabular}                & \multicolumn{1}{c|}{\begin{tabular}[c]{@{}c@{}}0.19\\ kg/hr\end{tabular}}                       & \begin{tabular}[c]{@{}c@{}}0.73\\ kg/hr\end{tabular}                           & \multicolumn{1}{c|}{\begin{tabular}[c]{@{}c@{}}-0.29\\ kg/hr\end{tabular}}              & \multicolumn{1}{c|}{\begin{tabular}[c]{@{}c@{}}0.72\\ kg/hr\end{tabular}}                     & \textbf{\begin{tabular}[c]{@{}c@{}}0.06\\ kg/hr\end{tabular}}     \\ \hline
\textbf{\begin{tabular}[c]{@{}c@{}}IQR of\\ site-level\\ quant errors\end{tabular}}                  & \multicolumn{1}{c|}{\begin{tabular}[c]{@{}c@{}}0.62\\ kg/hr\end{tabular}}               & \begin{tabular}[c]{@{}c@{}}0.57\\ kg/hr\end{tabular}                & \multicolumn{1}{c|}{\begin{tabular}[c]{@{}c@{}}0.61\\ kg/hr\end{tabular}}                       & \begin{tabular}[c]{@{}c@{}}4.28\\ kg/hr\end{tabular}                           & \multicolumn{1}{c|}{\begin{tabular}[c]{@{}c@{}}0.63\\ kg/hr\end{tabular}}               & \multicolumn{1}{c|}{\textbf{\begin{tabular}[c]{@{}c@{}}0.50\\ kg/hr\end{tabular}}}            & \begin{tabular}[c]{@{}c@{}}0.58\\ kg/hr\end{tabular}              \\ \hline
\textbf{\begin{tabular}[c]{@{}c@{}}Site-level \\ coverage\\ of 95\% CI\end{tabular}}                 & \multicolumn{1}{c|}{N/A}                                                                & N/A                                                                 & \multicolumn{1}{c|}{0.65}                                                                       & 0.37                                                                           & \multicolumn{1}{c|}{0.35}                                                               & \multicolumn{1}{c|}{0.55}                                                                     & \textbf{0.66}                                                     \\ \hline
\textbf{\begin{tabular}[c]{@{}c@{}}Avg number\\ of correct\\ localization \\ estimates\end{tabular}} & \multicolumn{1}{c|}{\textbf{\begin{tabular}[c]{@{}c@{}}4.14\\ out of\\ 5\end{tabular}}} & \begin{tabular}[c]{@{}c@{}}4.04\\ out of\\ 5\end{tabular}           & \multicolumn{1}{c|}{\begin{tabular}[c]{@{}c@{}}3.94\\ out of\\ 5\end{tabular}}                  & \begin{tabular}[c]{@{}c@{}}3.53\\ out of\\ 5\end{tabular}                      & \multicolumn{1}{c|}{\begin{tabular}[c]{@{}c@{}}3.97\\ out of\\ 5\end{tabular}}          & \multicolumn{1}{c|}{\begin{tabular}[c]{@{}c@{}}4.08\\ out of\\ 5\end{tabular}}                & \begin{tabular}[c]{@{}c@{}}4.01\\ out of\\ 5\end{tabular}         \\ \hline
\end{tabular}
\label{tab:comparison}
\end{table}

\section{Discussion}\label{sec:discussion}

We have proposed a Bayesian hierarchical model for methane emission source apportionment on oil and gas sites, referred to as the MDLQ model. This model is able to estimate total emissions on a representative oil and gas site to within 0.11 metric tons over the course of a three month experiment, a percent error of just 4.5\%.

The primary statistical contribution of this work is the synthesis and extension of other source apportionment methods from the existing literature, such as the spike-and-slab prior from \citet{weidmann_locating_2022}, the autocorrelated error model from \citet{ganesan_characterization_2014}, and the time-varying Gaussian puff atmospheric dispersion model from \citet{jia_fast_2025}. We extend these studies and synthesize their contributions in the MDLQ model. 

The contribution of this work to the field of methane emissions mitigation is twofold. First, we provide the most accurate method for constructing site-level emissions inventories with in situ sensors to date. Site-level inventories are an important tool for emissions accounting and mitigation in US regulations \citep{us_environmental_protection_agency_greenhouse_2024}, EU import standards \citep{EUreg}, and voluntary emission management protocols \citep{OGMP}. The inversion performance of many commercially available CMS solutions is lacking \citep{ilonze_assessing_2024, chen_comparing_2024, daniels_intercomparison_2025}, and we hope to elevate all solutions for this problem by making the MDLQ model open source. Second, the METEC experiment provides a more robust evaluation of source apportionment methods for oil and gas sites than other experiments in the literature, as it contains about 10 times more releases and uses more realistic emission characteristics. As such, we believe that the results in this paper are more likely to generalize to real oil and gas sites. 

We conclude by discussing several limitations of the MDLQ model and METEC experiment and how they may be addressed in future work. First, the MDLQ model does not estimate or adjust any parameters of the forward atmospheric dispersion model as is done in \citet{cartwright_bayesian_2019} or \citet{newman_probabilistic_2024} and instead prescribes them based on atmospheric stability class. This may contribute to the large variability in inferred emission rates, as the Pasquill-Gifford-Turner dispersion scheme can overestimate or underestimate true atmospheric dispersion. Developing a faster implementation of the Gaussian puff model (or another fast solution to the advection-diffusion equation) may make it feasible to sample dispersion parameters within the MCMC without having to use the lower fidelity Gaussian plume model.

Second, the METEC facility is designed to emulate relatively small oil and gas production sites with spatially distinct equipment groups. Other oil and gas facilities, such as midstream compressor stations, can be much larger with centrally housed equipment and overlapping emission sources. The MDLQ model may be able to apportion emissions between broad segments of these sites rather than specific equipment groups, but more testing is needed to assess this potential use case. More broadly, the MDLQ model is currently focused on apportionment problems where the number and location of potential sources is known. Oil and gas production sites are an important example of this type of apportionment problem, but they exist in other application domains as well, such as engineered landfills and carbon capture and storage facilities. Future work will extend the MDLQ model to apportionment problems where the number of potential sources is unknown. For example, the MDLQ model could be used to regularize a grid of emission sources using the spike-and-slab prior, similar to the method in \citet{weidmann_locating_2022}, or it could be used to refine a coarser estimate of source location. For the latter, the single-source localization procedure from \citet{daniels_detection_2024} could be iteratively applied to a grid of emission sources to identify the subset whose emissions (when forward simulated) exhibit strong correlation with observed concentrations; the MDLQ model could then further refine this subset.


Third, errors in the source-level inventories from the MDLQ model range from -37.7\% to 81.8\% when using all available data, but are significantly reduced when using just the inversion windows that have at least one sensor downwind of all sources for the entire window. With this in mind, we will investigate an adaptive windowing approach where the exact boundaries of each window are based on the information content of the CMS data rather than being fixed at 30-minute intervals. Additionally, using a probabilistic sampling scheme similar to the one proposed in \citet{daniels_estimating_2024} may provide a better way to extrapolate emission rate estimates from windows with full information to the desired temporal range (e.g., an annual inventory for regulatory reporting). We believe that these efforts would improve source-level inventories and alerts derived from the MDLQ model.

\section{Significance Statement} 

Efforts to reduce methane emissions from the oil and gas sector are increasingly based on measurements of ambient methane concentrations, but using these data to apportion emissions between sources on individual sites remains a challenge. This study pairs a new method for source apportionment with a controlled release dataset for model evaluation that accurately mimics real-world emission characteristics. By accounting for emission intermittency and autocorrelation through time, the model demonstrates state-of-the-art apportionment accuracy and uncertainty quantification. By making the model open-source, it can be adopted by oil and gas companies to prioritize their mitigation efforts or by regulators to verify their long-term emission reduction requirements. More broadly, the model is applicable to other source apportionment problems where the number and location of potential sources are known, extending its relevance beyond oil and gas methane emissions.

\begin{acks}[Acknowledgments]
The authors would like to thank Cal Okenberg for identifying the \texttt{TruncatedNormal} package in \texttt{R} for use in the MCMC sampler and for reviewing the manuscript. The authors would also like to thank Troy Sorensen and Spencer Kidd for testing the MDLQ algorithm on multiple oil and gas sites and for reviewing the manuscript.
\end{acks}
\bibliographystyle{imsart-nameyear} 
\bibliography{references}       


\end{document}